\documentclass[12pt]{article}
\usepackage{epsfig, amssymb, graphicx, amsmath} 
\usepackage{amsthm}
\usepackage{relsize}
\usepackage{comment}

\usepackage[round-mode=places, round-integer-to-decimal, round-precision=2,
table-format = 1.2, 
table-number-alignment=center,
round-integer-to-decimal,
output-decimal-marker={,}
]{siunitx} 
\usepackage{booktabs}
\usepackage{enumitem}
\usepackage{eurosym}	
\usepackage{wrapfig}	
\usepackage{float}		
\usepackage{footnote}
\usepackage{amssymb}
\usepackage{bm}
\usepackage{float}
\usepackage{booktabs}
\usepackage{xcolor}
\usepackage{colortbl}
\usepackage{graphbox}

\usepackage{comment}
\usepackage{rotating}

\usepackage{thmtools}
\usepackage{bbm}		
\usepackage[title]{appendix}
\usepackage{adjustbox}
\usepackage{multirow}
\usepackage{hyperref}
\usepackage{multicol}
\usepackage{subfig}	
\usepackage[authoryear]{natbib}
\bibpunct{(}{)}{,}{a}{}{,}

\numberwithin{equation}{section}

\theoremstyle{definition}

\theoremstyle{remark}

\numberwithin{theorem}{section}
\numberwithin{proposition}{section}
\numberwithin{lemma}{section}
\numberwithin{corollary}{section}
\numberwithin{definition}{section}
\numberwithin{remark}{section}
\numberwithin{example}{section}

\newcommand{\be}{\begin{equation}}
	\newcommand{\en}{\end{equation}}
\newcommand{\ben}{\begin{equation*}}
	\newcommand{\enn}{\end{equation*}}
\newcommand{\bea}{\begin{eqnarray}}
	\newcommand{\ena}{\end{eqnarray}}

\begin{document}
	
	\newlength\tindent
	\setlength{\tindent}{\parindent}
	\setlength{\parindent}{0pt}
	\renewcommand{\indent}{\hspace*{\tindent}}
	
	\begin{savenotes}
		\title{
			\bf{ 
				Hedging carbon risk with a network approach
		}}
		\author{
			Michele Azzone$^*$ ,
			Maria Chiara Pocelli$^*$ ,
			Davide Stocco$^*$
		}
		
		\maketitle
		
		\vspace*{0.11truein}
		\begin{tabular}{ll}
			$*$ & Politecnico di Milano, Department of Mathematics, Italy.\\
			
		\end{tabular}
	\end{savenotes}
	
	\vspace*{0.11truein}
\begin{abstract}
Sustainable investing refers to the integration of environmental and social aspects in investors' decisions.  We propose a novel methodology based on the Triangulated Maximally Filtered Graph and node2vec algorithms to construct an hedging portfolio for climate risk, represented by various  risk factors, among which the $CO_2$ and the ESG ones.
The $CO_2$ factor is strongly correlated consistently over time with the Utility sector,  which is the most carbon intensive in the S\&P 500 index. Conversely, identifying a group of sectors linked to the  ESG factor proves challenging. As a consequence, while it is possible to obtain an efficient hedging portfolio strategy with our methodology for the carbon factor, the same cannot be achieved for the ESG one. The ESG scores appears to be an indicator too broadly defined for market applications.  These results support the idea  that bank capital requirements should take into account  carbon risk.
\end{abstract}

	\vspace*{0.11truein}
	\vspace*{0.11truein}

\section{Introduction}
\label{}
After the Paris Agremeent, green and sustainable investments have been in the spotlight \citep{fahmy2022rise}.
Understanding the interplay between this non-traditional financial information and  asset prices appears to be crucial both from a practitioner and academic perspective to enhance sustainable investing. 
The relationship between ESG ratings and stocks returns has been vastly investigated in recent literature. 

However, the empirical studies show heterogeneous results on the influence of ESG ratings on the market, mostly driven by disagreement in the scores provided by the rating agencies \citep{berg2022aggregate, billio2021inside, gibson2021esg}. 
\smallskip

A substantial number of studies investigate the presence of an ESG-related market factor and its impact on investments.
\citet{lioui2022chasing} propose a factor construction methodology that controls for ESG ratings and other firm characteristics. They  find conflicting evidence regarding the ESG factor $\alpha$ which changes sign over time.
\citet{naffa2022factor} find no sufficient evidence for ESG factors to complement
the \cite{fama2015five} 5 factor models (FF5). 
They find that ESG portfolios (factors) do not generate significant alphas during 2015-2019.  
\cite{bang2023esg} compare the ESG and the ESG controversy risks: their study shows that the ESG factor itself is an idiosyncratic risk, while the controversy factor has additional explanatory power for stock returns.
\smallskip

It follows from the discussion above that  no general consensus has been achieved on the existence of an ESG pricing factor. For this reason the attention has been moved towards different sustainability-related information that could unveil the nexus between this new set of information and the market behaviour.
\cite{bams2022tilting} split ESG information into two subsets, namely ESG promised and realized data, creating two different ratings. They show that firms are pushed to inflate their ESG ratings to lower capital costs. \citet{agliardi2023environmental} instead claim that the information contained in the Environmental pillar is influencing stock returns and that environmentally low-rated companies present better financial performances.
\smallskip

Another stream of literature focuses on the solely $CO_2$ emissions and the so-called carbon premium. \cite{bolton2021investors} show that the carbon premium cannot be explained by known risk factors. 
This result reinforces the idea that
the level of carbon emissions contains independent information about the cross-section of average returns. 
Moreover, they show that high polluting companies are associated to higher returns because investors are already aware of their exposure to carbon transition risk, demanding further compensation for it \citep[see e.g.,][]{freire2019reformulating,flora2023green}.
These extra profits appear to be lowered during periods of high attention towards climate news \citep{huij2022carbon, liu2021relationship}.
\smallskip

One of the holy grails in the field is finding evidence of this carbon premium or carbon risk factor, which appear to be crucial both in developed and medium- and low-income countries investments \citet{wang2024have}.
\citet{batten2018addressing} proposed an asset management technique for a portfolio that provides risk-adjusted positive benefits to investors, while  changing weights over time as COP21 implementation proceeds.
\citet{gorgen2020carbon} do not find evidence of a carbon risk premium, using a factor-mimicking portfolio approach. 
\citet{witkowski2021does} investigate if there is a stable carbon premium in energy-intensive sectors. They find statistically significant results only for some time intervals. \citet{huij2022carbon} introduce a solid methodology to compute a $CO_2$ risk factor which, they show, is linked to stock returns.

\smallskip

An important part of a bank’s trading desk risk management  is the identification of a proper set of market risk factors. The risk factors contained in a bank's model must be sufficient to represent the risks inherent in the underlying portfolio. In particular, for equity prices, there must be a set of  risk factors for every equity markets in which the bank holds significant positions. At least, there must be the risk factor corresponding to  market-wide movements (e.g. the CAPM market risk factor). However, the set of considered factors for a given market should
correspond to the bank’s exposure to the overall market as well as its concentration in
individual stocks in that market.\footnote{See the Basel Committee on Banking Supervision Calculation of RWA for market risk \url{https://www.bis.org/basel_framework/chapter/MAR/31.htm?inforce=20230101&published=20200605} and the Minimum Capital Requirements for market risk \url{https://www.bis.org/bcbs/publ/d352.pdf}}
\smallskip

Hedging risk factors plays a key role in determining the bank capital requirements because the business model of trading desks is usually  built upon hedging most of the risky positions.\footnote{Commercial banks have been forbidden speculative proprietary trading operations after the Great Financial Crisis \url{https://www.bis.org/publ/cgfs52.pdf}}
It is well established that the standard CAPM market risk factor can be hedged considering long-short positions and the corresponding betas (this procedure is detailed in the BIS Minimum Capital Requirements for market risk \url{https://www.bis.org/bcbs/publ/d352.pdf}). Similar approach can be followed to hedge the FF5 factors. However, there is still lack of knowledge on the relationship between the carbon risk factor and energy intensive stocks, and the possible related hedging strategies.
Similar conclusions can be drawn for a potential ESG market risk factor \citep{pollard2018establishing, cornell2021esg}.
\smallskip

Our analysis lies in this blind spot. We propose a novel methodology based on a complex network market representation coupled with a low dimensional embedding technique to analyze the significance of ESG and carbon risks in the market. We build four risk factors, namely ESG, ESG promised, ESG realized and $CO_2$,  following the same approach that  \citet{huij2022carbon} employed for the $CO_2$ factor.
Specifically, we utilize the  Triangulated Maximally Filtered Graph (TMFG) introduced by \citet{massara2016network} to build a sparsely connected graph from the correlation matrix of S\&P 500 stocks' returns.
Our key proposal is then to  utilize the node2vec methodology \citep{grover2016node2vec}, a representational learning technique, to project this graph into a low-dimensional space. Thanks to the natural distance defined in the embedded space, we can study whether a risk factor is close to individual stocks or sectors and its stability over time.  We  define an hedging portfolio that relies only on a limited number of long and short position in the S\&P 500 stocks.  Employing this simple strategy, we are able to identify  hedging portfolios that have low exposure to the market and are negatively correlated with a specific factor.
\smallskip

Three are the main result of this paper. First,
we introduce a novel methodology to construct hedging portfolios to any risk factor, combining the TGMF algorithm with the node2vec embedding technique.
Second, we apply our method to a set of sustainability related risk factors.
We observe that the $CO_2$ factor is stable over time and is always close (in the sense of the embedded space metric) to the Utility sector,  which is the most carbon intensive in the S\&P 500 index. However, similar results do not hold for the three ESG factors. Third, it is possible to obtain an hedging portfolio (negatively correlated with the factor)  for each factor except the ESG one. The latter seems to be an indicator too nebulous to be hedged efficiently.  It is worth noting that the $CO_2$ hedge portfolio is short the Utility sector. We investigate the performances and the risk of the $CO_2$ hedge portfolio and verify that it is not exposed to consistent losses in terms of standard risk metrics.

The rest of the paper is organised as follows.
In Section \ref{methodology_dataset}, we discuss the methodology and the dataset used in this work.
In Section \ref{results}, we present the main results on the set of sustainability risk factors.
In Section \ref{conclusion}, we draw the conclusions of the analysis.  Finally, in the appendix, we report  additional numerical results.


\section{How to hedge a riks factor}
\label{methodology_dataset}
\subsection{Methodology}
\label{methodology}
In this section, we illustrate the methodology that we utilize to build the four risk factors and  describe the network analysis.
\smallskip

In our analysis, we consider four different measures of sustainability: the $CO_2$ total emissions, the ESG scores (from  Refinitiv), the ESG promised and the ESG realized scores built starting from granular data. Following \cite{huij2022carbon}, we calculate the CO2 total emissions as the sum of $CO_2$ Scope 1 (direct emissions) and Scope 2 (indirect emissions) divided by the company's market capitalization, to account for firm size. We use this weighted emission score to build the $CO_2$ factor.
Following the approach proposed by \cite{bams2022tilting}, we create two subsets of ESG granular information.
We segregate ESG activities, policies, reporting and targets from the ESG controversies and performances granular data, to create respectively the promised and the realized ESG scores for each firm. 
We are then able to build the $CO_2$, ESG, ESG promised and ESG realized factors with respect to the scores mentioned above.
Each factor is created using a long-short position with a similar approach to the Fama-French HML portfolio construction \citep{fama1993common}. 
Specifically, each factor is an equally-weighted portfolio that takes long (short) position in the 30\% firms achieving the highest (lowest) values of the corresponding measures.
To be as general as possible in our analysis, we construct the $CO_2$ and ESG factors on the set of Russell 3000  stocks, and the ESG promised and realized factors on the S\&P 500 stocks, due to the lack of accurate ESG granular data on small cap. We include such factors in our study as proxies for sustainability risks. 
\smallskip

We use a complex network methodology to study the relationship between our factors and stocks and sectors in the financial market. We proceed to the construction of a financial graph, where nodes represent stocks and edges between nodes are weighted according to a similarity measure defined below. We also add four fictitious nodes to the graph, namely the four factors introduced above.  
To determine the edges' weights, we first estimate the FF5 linear regression to the daily logarithmic returns of the stocks. We then take into account only the residual returns, to remove the uninformative common behaviour of stocks described by the FF5 risk factors. We finally use the Pearson correlation measure \citep[see][]{cohen2009pearson} to estimate the relationship between stocks' residual returns and factors and consider the correlation matrix as the weighted adjacency matrix of the financial graph. 
\smallskip

By construction, the correlation graph results to be fully connected, including uninformative connections.
We thus employ the Triangulated Maximally Filtered Graph (TMFG) tool \citep{massara2016network}, to reduce complexity while preserving relevant data structure. Specifically, the TMFG efficiently builds a planar filtered graph, where only relevant edges are preserved (filtered graph), and edge-crossing is avoided (planarity constraint). We refer to \citet{massara2016network} for a comprehensive discussion of the TMFG methodology and its advantages.
We point out that this network representation is characterized by relational bonds between different nodes. In this framework, it is clearly possible to define different distances on the network to understand which node is ``close" or ``far" to another (e.g. if a stock is close or far from a factor).

Instead, we consider a different approach. Our proposal is to embed the filtered graph in a low dimensional space using the node2vec method  \citep{grover2016node2vec}. This algorithm projects a graph into a low-dimensional vector space,  where each node corresponds to a point, while maximizing the amount of retained relational information. In this way, the analysis of the graph is greatly simplified, as relationships among nodes can be quantified by employing common distance metrics. By applying the node2vec algorithm to our filtered network we end-up with a lower dimensional representation of the relationships between  nodes while minimizing the information loss. For these reasons, we claim that the node2vec algorithm is a viable and robust choice also from a practitioner's perspective. Up to our knowledge, this is the first application of the node2vec algorithm in the financial literature.
\smallskip

In the embedded vector space obtained from the node2vec algorithm, we employ  the standard Euclidean distance  to build  effective investment portfolios. We repeat the procedure described below for each risk factor considered in the study. We first take into account the 30 closest firms to a given factor and construct the \textit{close} portfolio. In the same way, we construct the \textit{far} portfolio, composed of the 30 furthest firms from such factor. We then compare the performances of the two portfolios with the ones of the \textit{far-close} portfolio, consisting of a long (short) position in the top 15 firms far from (close to) the factor. All the portfolios are equally weighted. We calculate their daily logarithmic returns and study their exposure to the FF5 factors and to the additional sustainability factor. Moreover, we analyze their performances in terms of Sharpe ratio, Sortino ratio, Omega ratio, Maximum DrawDown (MDD), and  5\% Value at Risk (VaR); for a detailed description of these metrics \citep[see e.g.,][]{billio2021inside}.

\subsection{Dataset}
\label{dataset}
Our dataset spans from $1^{st}$ January 2015 to $31^{st}$ December 2020 and is sourced from Refinitiv Eikon Reuters.
The selected time interval goes from the Paris Agreement adoption in 2015 to the U.S. withdrawal from it in 2020. We consider stocks in the Russell 3000 and the S\&P 500 Index. We use the stocks from the Russell 3000 to build the $CO_2$ and the ESG factor and, due to the lack of accurate ESG granular data, stocks from the S\&P 500 Index to create the ESG promised and realized factors.
\smallskip

We select only stocks that remained part of the S\&P 500 Index throughout this entire six-year period to construct our financial network. The resulting dataset is composed of $470$ stocks, each of them described by a vector of daily price observations. These stocks are categorized into $11$ sectors according to the Global Industry Classification Standard (GICS), as summarized in Table \ref{table:gics}. Among these sectors, Financials, Industrials, Health Care,  Communication Services and Information Technology  represent the highest percentages of market capitalization.
\begin{table}[H]
	\scriptsize
	\centering 
	\begin{tabular}{l c c c c c c}
		\toprule
		\multirow{ 2}{*}{\textbf{GICS Sector}} & \multirow{ 2}{*}{\textbf{\#Stocks}} & \textbf{Market Cap}&\textbf{$\mathbf{CO_2}$}&  \multirow{ 2}{*}{\textbf{ESG}} & \multirow{ 2}{*}{\textbf{ESGp}} & \multirow{ 2}{*}{\textbf{ESGr}}\\
		& & \textbf{[trillion\$]} & \bm{$[10^{5}mt/\$]$} & & & \\
		\midrule
		\rowcolor{gray!15}
		Financials & 71 & 3.1 (15.7\%)&0.3 & 53.3 & 56.2 &	49.3 \\  
		Industrials & 69 & 2.0 (10.3\%) &23.3 & 56.6 & 56.7 & 49.6\\
		\rowcolor{gray!15}
		Health Care & 61 & 2.6 (13.4\%) & 0.8 & 58.3 & 56.8 & 50.4 \\
		Information Technology & 61 & 3.0(15.6\%) & 2.2 & 59.1 & 58.2 & 52.2 \\
		\rowcolor{gray!15}
		Consumer Discretionary & 51 & 1.8 (9.1\%) & 8.4 & 53.5 & 56.7 & 50.2 \\
		Consumer Staples & 34 & 1.9 (9.6\%)  & 4.9 & 65.9 & 57.1 & 51.1 \\
		\rowcolor{gray!15}
		Utilities & 29 & 0.6 (3\%) & 179.5 & 59.9 & 56.8 & 50.4 \\
		Real Estate & 27 & 0.5 (2.6\%) & 2.7 & 59.3 & 56.8 & 50.2 \\
		\rowcolor{gray!15}
		Energy & 23 & 1.2 (6.3\%) & 35.0 & 58.6 & 56.8 & 50.4 \\
		Materials & 23 & 0.4 (1.9\%) & 30.6 & 61.5 & 56.7 & 50.3 \\
		\rowcolor{gray!15}
		Communication Services & 21 & 2.5 (12.4\%)& 3.0 & 46.5 & 56.7 & 50.5 \\
		\midrule
		\textbf{Total} & \textbf{470} & \textbf{19.8}&-&-&-&- \\
		\bottomrule
	\end{tabular}
	\\[10pt]
	\caption{Number of stocks and average market capitalization, $CO_2$ total emission, ESG, ESG promised, and ESG realized scores for every GICS sector.}
	\label{table:gics}
\end{table}
In Table \ref{table:gics}, we also report the averages of $CO_2$ total emissions, ESG, ESG promised and ESG realized scores by sector. There is a clear difference between sectors in terms of $CO_2$ emissions, with Utilities displaying notably the highest values. Let us emphasize that the same does not hold for the three ESG-related factors which are less variable across sectors.


\section{Results}\label{results}
In this section, we report the main results of the application of our methodology on the four sustainability-related risk factors and the S\&P500 stocks. We show that the $CO_2$ factor is close (for different time windows) to the Utility sector while there is no clear connection between the position of the other three ESG-based factors and the different sectors. Moreover, we discuss how it is possible to construct efficient hedging portfolios for all factors except the ESG one. Finally, we show that the $CO_2$ hedging portfolio has low exposure to market losses in terms of standard risk metrics.
\smallskip

\begin{figure}[H]
	\centering
	\subfloat[Planar filtered graph. Edge width is proportional to the corresponding edge weight.\label{fig:tmfg_6y}]{
		\includegraphics[scale=0.2]{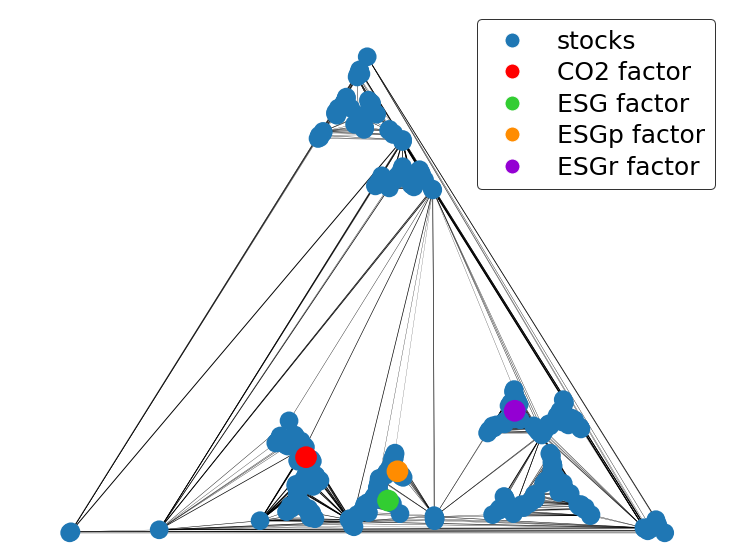}
	}
	\hfill
	\subfloat[2D map of the learned embedded space. Node colors reflect the GICS sectors; fictitious nodes - in red - are labelled.\label{fig:node2vec_6y}]{
		\includegraphics[scale=0.19]{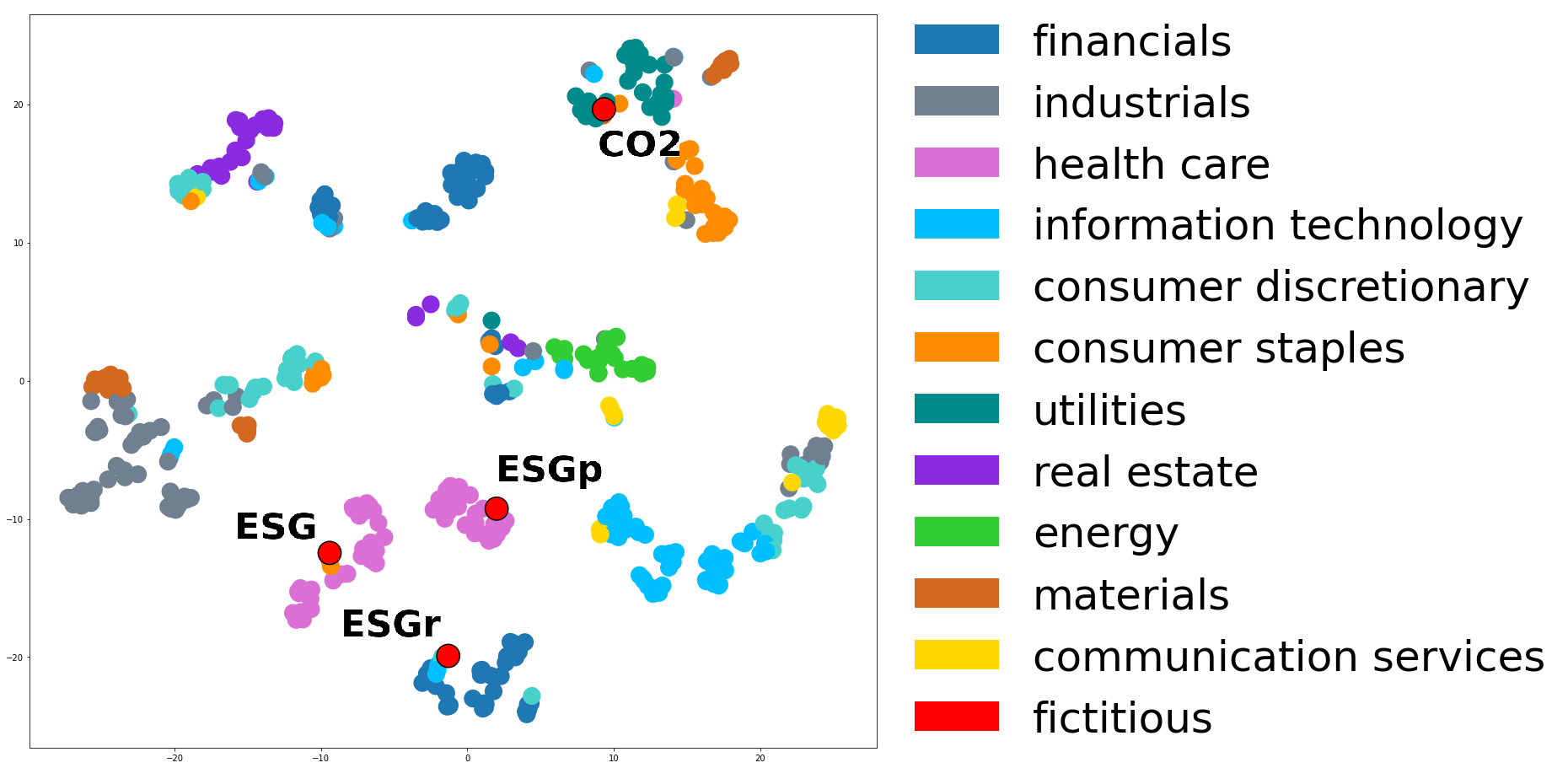}
	}
	\caption{2015-2020 time period.}
	\label{fig:6y}
\end{figure}

As discussed in the previous section, we start our analysis by constructing the financial graph  on the correlation matrix of the four proposed factors and the S\&P 500 stocks in the 2015-2020 time window. We apply the TMFG algorithm to trim the uninformative edges of the graph and utilize the node2vec procedure to obtain a  visual representation of the sparsely connected network in the embedded low dimensional space. 
In Figure \ref{fig:tmfg_6y}, we report a visual representation of the sparsely connected network for the entire time-interval of analysis. We can observe how the resulting network is composed solely of cliques. Let us notice that the visual representation is  not informative of the distances between nodes and cannot be used to infer the relationships among stocks and factors we are interested in. As discussed in section \ref{dataset}, we employ node2vec to solve this issue. 
\smallskip

\begin{figure}[H]
	\centering
	\includegraphics[scale=0.3]{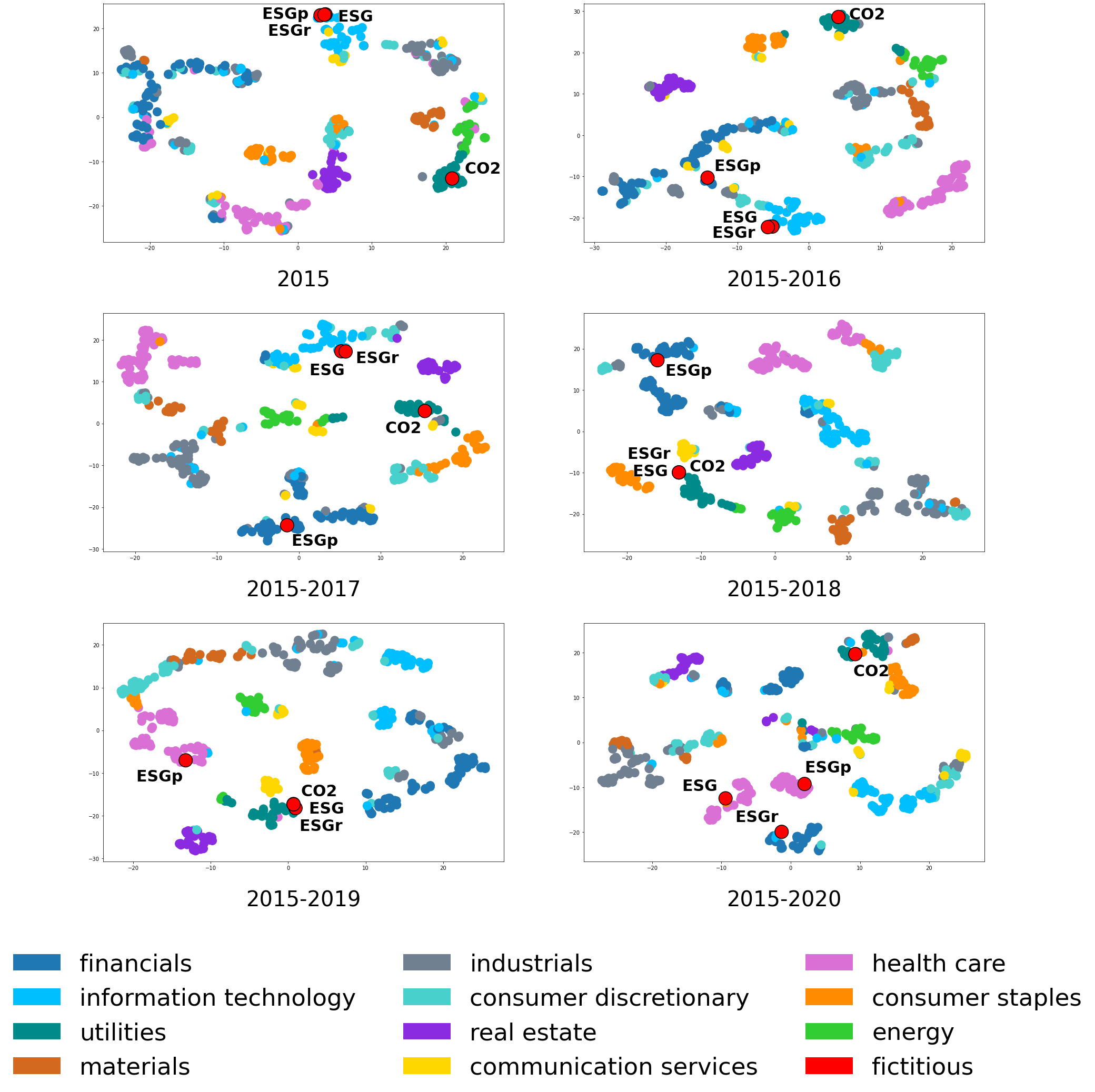}
	\caption{2D map of the learned embedded space for an expanding window starting from 2015. Node colors reflect the GICS sectors; fictitious nodes - in red - are labelled. The $CO_2$ factor is consistently close to the Utility sector while the three ESG factor are not close to a specific sector in different  time-windows.}
	\label{fig:node2vecannual}
\end{figure}

Figure \ref{fig:node2vec_6y} displays the  2D map of the  embedded space obtained with the node2vec algorithm.
Within this space, defining a standard Euclidean metric  is elementary. Several noteworthy observations can be made. First, let us observe that the different sectors ( different sectors are identified by the colour of the dots) appear clearly separated. Second,  we note that $CO_2$ factor is located in the center of the Utility sector. This is a particularly  interesting result, considering that this sector is the most carbon intensive (cf. Table \ref{table:gics}). Finally, we point out that it is more difficult to interpret the relationship between the  three ESG factors and the S\&P 500 stocks. The standard ESG factor  and the promised ESG factor appear close to the Health-Care sector which exhibits average performances in terms of ESG and ESG promised. The ESG realized factor is instead  close to the Financial sector which has the lowest ESG  realized score.
We can conclude that in the 2015-2020 time window the $CO_2$ factor  is in close proximity to a cluster of stocks with high carbon emission (that corresponds to the Utility sector), while the three ESG factor positions pose a more complex challenge in terms of interpretation. In Figure \ref{fig:node2vecannual}, we  delve deeper into the consistency  of  the position of these factors by considering an extending training window for the TMFG and the node2vec algorithms (from 2015 to the range 2015-2020).
Throughout the considered time windows, the $CO_2$ factor is consistently close to the Utility sector, while the three ESG factors are not close to the same sector in different  time windows (e.g. they are close to Information Technologies in 2015-2016 and Health-Care in 2015-2020). Additionally, the different ESG factors are close in some time windows and spread out in some others. For instance, they are in close proximity in 2015 time window and spread out in the 2015-2016 period. These results further confirm the reliability of the $CO_2$ factor and its strong connection with the highly polluting stocks. Unfortunately, the same cannot be stated for the three ESG related risk factors.

As discussed in section \ref{methodology}, we  construct the \textit{close}, \textit{far} and \textit{far-close} portfolios for each factor. We want to verify whether these portfolios are correlated with the factors and evaluate their performances. In the following, we present our results for the $CO_2$ factor which, as already observed, is closely linked to the Utility sector. To this aim, we regress the returns of the \textit{close}, \textit{far} and \textit{far-close} portfolios on the \citet{fama1993common} three factors (FF3) considering both the standard FF3 regression and the case when we add the $CO_2$ factor. We present several robustness regressions in the appendix which confirm the results presented in the rest of the section.

In Table \ref{table:ff4}, we report the results of the six regressions, considering Newey-West standard errors. On the left we show the results without the $CO_2$ factor and on the right the ones obtained adding the $CO_2$ factor. When we include the $CO_2$ factor in the regressions, the adjusted R-squared always increases. The $CO_2$ factor is significant at $5\%$ for the \textit{close} portfolio and the \textit{far-close} portfolio. The estimated coefficient is positive for the former and negative for the latter one.  Therefore, the $CO_2$ factor is a significant driver in the \textit{close} portfolio, which is highly exposed to $CO_2$ emissions because is mainly composed by Utility stocks, and can be hedged efficiently with a long-short position in green and brown stocks (i.e. our \textit{far-close} portfolio). Moreover, our results on the \textit{close} portfolio are consistent with an interpretation that investors are already pricing a compensation for their exposure to $CO_2$ risk. The same results are confirmed  when we repeat the analysis considering either the standard CAPM model or all the FF5  factors. The results of these additional analyses are shown in the appendix. Let us emphasize that the significance of the $CO_2$ factor for the considered portfolios goes in the direction of confirming the existence of a carbon risk factor which is not taken into account in the traditional FF3 or FF5 risk factors \citep[see e.g.,][]{bolton2021investors,huij2022carbon}.

\begin{table}[H]
	\scriptsize
	\centering 
	\begin{tabular}{l l l l l l l}
		\toprule
		& \multicolumn{3}{c}{\textbf{Three-factor}} & \multicolumn{3}{c}{\textbf{Three-factor+$CO_2$}} \\
		\cmidrule(r){2-4} \cmidrule(l){5-7}
		& \textbf{\textit{close}} & \textbf{\textit{far}} & \textbf{\textit{far-close}} & \textbf{\textit{close}} & \textbf{\textit{far}} & \textbf{\textit{far-close}} \\
		\midrule
		\rowcolor{gray!15} 
		\bm{$\alpha$} & -0.0001 & -0.0001 & -0.0000 & 0.0000 & -0.0001 & -0.0000 \\ 
		\rowcolor{gray!15} 
		& (0.0002) & (0.0002) & (0.0001) & (0.0002) & (0.0002) & (0.0001)\\ 
		\bm{$r_M-r_F$} & 0.7546*** & 1.0469*** & 0.1551*** & 0.7964*** & 1.0549*** & 0.1367*** \\  
		& (0.0423) & (0.0316) & (0.0212) & (0.0444) & (0.0318) & (0.0222)\\ 
		\rowcolor{gray!15} 
		\bm{$SMB$} & -0.1753*** & 0.2538*** & 0.2310*** & -0.2070*** & 0.2477*** & 0.2449*** \\
		\rowcolor{gray!15} 
		& (0.0506) & (0.0536) & (0.0328) & (0.0499) & (0.0535) & (0.0332)\\
		\bm{$HML$} & 0.1689*** & 0.1887*** & -0.0066 & 0.0166 & 0.1594*** & 0.0603 \\  
		& (0.0367) & (0.0307) & (0.026) & (0.0700) & (0.0365) & (0.0398)\\  
		\rowcolor{gray!15} 
		\bm{$CO_2$} & - & - & - & 0.3606** & 0.0694* & -0.1582** \\ 
		\rowcolor{gray!15}
		& - & - & - & (0.1475) & (0.0380) & (0.0709)\\ 
		\textbf{R-squared} & 0.5773 & 0.8595 & 0.1971 & 0.6195 & 0.8606 & 0.2330 \\
		\bottomrule
	\end{tabular}
	\\[10pt]
	\caption{Fama-French three-factor and three-factor+$CO_2$ regressions output for the \textit{close}, \textit{far} and \textit{far-close} portfolios. Statistical significance at the $10\%$, $5\%$ and $1\%$ is denoted by *, ** and ***, respectively. HAC robust standard errors are indicated in brackets.}
	\label{table:ff4}
\end{table}

We have just described the relationship between the $CO_2$ risk factor and some relevant portfolios of S\&P 500 stocks. We claim that the \textit{close} portfolio is exposed to carbon risk while an hedging opportunity is embodied by the \textit{far-close} one. A natural question arises: what are the performances of such portfolios?
\smallskip

In Table \ref{table:performance}, we report the Sharpe, Sortino and Omega  ratio for the three portfolios of interest to evaluate and compare their return and risk profiles against two benchmark: the \textit{S\&P} portfolio, composed of all 470 stocks of our dataset with equal weighting, and a \textit{random} portfolio, composed of 30 randomly selected stocks from our dataset. We report also the MDD and 5\% VaR. The \textit{close} and the \textit{far} portfolio outperforms the \textit{far-close} portfolio in terms of all ratios (the \textit{far} portfolio being the most efficient). Interestingly, the green \textit{far} portfolio outperforms also the \textit{S\&P} portfolio in the period under analysis. This is in agreement with  \citet{PERERA2023107053} that show how brown firm returns present an higher variance (and consequently a lower Sharpe ratio) than their greener counterparts. However, the MDD and VaR of the \textit{far-close} portfolio is significantly lower than the alternative ones. This supports our claim that the \textit{far-close} portfolio is an efficient hedging opportunity which provides a negative exposure to $CO_2$ while ensuring low correlation with market risk. Similar results hold also when analyzing the performances of portfolios constructed for the ESG factors (see Table \ref{table:performance_factors}).
\begin{table}[H]
	\scriptsize
	\centering 
	\begin{tabular}{p{10em} c c c c c}
		\toprule
		& \textbf{\textit{close}} & \textbf{\textit{far}} & \textbf{\textit{far-close}} & \textbf{\textit{S\&P}} & \textbf{\textit{random}} \\
		\midrule
		\rowcolor{gray!15}
		\textbf{Sharpe ratio} & 0.373 & 0.472 & 0.212 & 0.415 & 0.374 \\ 
		\textbf{Sortino ratio} & 0.423 & 0.497 & 0.288 & 0.452 & 0.408 \\ 
		\rowcolor{gray!15}
		\textbf{Omega ratio} & 1.091 & 1.116 & 1.056 & 1.102 & 1.091 \\ 
		\textbf{MDD} & 38.9\% & 50.0\% & 24.8\% & 42.9\% & 47.1\% \\ 
		\rowcolor{gray!15}
		\textbf{VaR 5\%} & 1.9\% & 2.3\% & 0.9\% & 2.0\% & 2.2\% \\ 
		\bottomrule
	\end{tabular}
	\\[10pt]
	\caption{Evaluation measures for the \textit{close}, \textit{far}, \textit{far-close}, \textit{S\&P}, and \textit{random} portfolios.}
	\label{table:performance}
\end{table}

The $CO_2$ appears closely linked to the Utility sector and we show that it can be hedged with a simple long-short position. Unfortunately, we do not achieve similar results when we consider the ESG factor. We repeat the analysis for the ESG, ESG promised and ESG realized factors. For each of them we build the \textit{close}, \textit{far} and \textit{far-close} portfolio, and for each portfolio we estimate a FF3 model with the additional factor as regressors. The results are available in Table \ref{table:ff4_factors}. For the ESG promised and the ESG realized cases, the results are consistent with the $CO_2$ case (even if the improvement in terms of  R-squared when adding the corresponding factor to the FF3 is, on average, inferior, see Table \ref{table:R2}). In both cases, the \textit{close} portfolio is positively correlated with the factor (ESG promised and realized) and the \textit{far-close} is negatively correlated. Instead, for the plain ESG case there is no statistical evidence that the hedging \textit{far-close} portfolios is negatively correlated with the ESG factor. These results confirm what has been pointed out by \citet{bams2022tilting}. The ESG factor is too broad an indicator to be used to constract a meaningful risk factor.  However, if we split the indicators used in its construction into more focused subsets, in our case the ESG promised and realized, we are able to identify significant risk factors. Let us point out, that also $CO_2$ emissions are one of the indicators that contributes to the formulation of the ESG score and thus, also the carbon score can be viewed as another sub-indicator of the ESG one.
\begin{table}[H]
	\centering 
	\resizebox{\textwidth}{!}{  \begin{tabular}{l l l l l l l l l l}
			\toprule
			& \multicolumn{3}{c}{\textbf{Three-factor+ESG}} & \multicolumn{3}{c}{\textbf{Three-factor+ESGp}} & \multicolumn{3}{c}{\textbf{Three-factor+ESGr}}\\
			\cmidrule(r){2-4} \cmidrule(l){5-7} \cmidrule(l){8-10}
			& \textbf{\textit{close}} & \textbf{\textit{far}} & \textbf{\textit{far-close}} & \textbf{\textit{close}} & \textbf{\textit{far}} & \textbf{\textit{far-close}} & \textbf{\textit{close}} & \textbf{\textit{far}} & \textbf{\textit{far-close}}\\
			\midrule
			\rowcolor{gray!15} 
			\bm{$\alpha$} & -0.0001 & -0.0002 & 0.0001 & 0.0001 & -0.0001 & -0.0001 & -0.0000 & -0.0001 & -0.0000 \\  
			\rowcolor{gray!15}
			& (0.0001) & (0.0002) & (0.0001) & (0.0001) & (0.0002) & (0.0001) & (0.0001) & (0.0002) & (0.0001) \\ 
			\bm{$r_M-r_F$} & 0.9338*** & 1.0424*** & 0.0685** & 0.9719*** & 0.9789*** & 0.0547** & 1.0364*** & 0.9877*** & -0.0844** \\ 
			& (0.0118) & (0.0425) & (0.0297) & (0.0150) & (0.0368) & (0.0248) & (0.0155) & (0.0422) & (0.0361) \\  
			\rowcolor{gray!15}
			\bm{$SMB$} & 0.1060*** & 0.2401*** & 0.0484 & 0.1272*** & 0.1238** & -0.0004 & 0.2019*** & 0.1391* & -0.0860 \\  
			\rowcolor{gray!15}
			& (0.0158) & (0.0794) & (0.0561) & (0.0316) & (0.0623) & (0.0356) & (0.0255) & (0.0837) & (0.0606)\\  
			\bm{$HML$} & -0.1791*** & 0.4197*** & 0.3117*** & -0.3302*** & 0.3679*** & 0.3321*** & 0.0736*** & 0.3427*** & 0.1714*** \\ 
			& (0.0149) & (0.0442) & (0.0284) & (0.0205) & (0.0363) & (0.0204) & (0.0179) & (0.0360) & (0.0260) \\ 
			\rowcolor{gray!15}
			\bm{$factor$} & -0.0415*** & -0.0527 & -0.0237 & 0.0603*** & -0.0725 & -0.0830** & 0.0271*** & -0.1292 & -0.0752* \\ 
			\rowcolor{gray!15}
			& (0.0157) & (0.0483) & (0.0316) & (0.0214) & (0.0551) & (0.0328) & (0.0096) & (0.0802) & (0.0440) \\ 
			\textbf{R-squared} & 0.9138 & 0.8005 & 0.3024 & 0.8785 & 0.8235 & 0.3423 & 0.9203 & 0.8124 & 0.1021 \\ 
			\bottomrule
	\end{tabular}}
	\\[10pt]
	\caption{Fama-French three-factor+ESG/ESGp/ESGr regressions output for the \textit{close}, \textit{far} and \textit{far-close} portfolios. Statistical significance at the $10\%$, $5\%$ and $1\%$ is denoted by *, ** and ***, respectively. HAC robust standard errors are indicated in brackets.}
	\label{table:ff4_factors}
\end{table}

\begin{table}[H]
	\scriptsize
	\centering 
	\begin{tabular}{p{5em} c c c c c c}
		\toprule
		& \multicolumn{3}{c}{\textbf{Three-factor}} & \multicolumn{3}{c}{\textbf{Three-factor+}\bm{$factor$}} \\
		\cmidrule(r){2-4} \cmidrule(l){5-7}
		& \textbf{\textit{close}} & \textbf{\textit{far}} & \textbf{\textit{far-close}} & \textbf{\textit{close}} & \textbf{\textit{far}} & \textbf{\textit{far-close}} \\
		\midrule
		\rowcolor{gray!15}
		\textbf{ESG} & 0.9135 & 0.8003 & 0.3020 & 0.9138 & 0.8005 & 0.3024 \\
		\textbf{ESGp} & 0.8763 & 0.8210 & 0.3181 & 0.8785 & 0.8235 & 0.3423 \\
		\rowcolor{gray!15}
		\textbf{ESGr} & 0.9201 & 0.8073 & 0.0899 & 0.9203 & 0.8124 & 0.1021 \\
		\bottomrule
	\end{tabular}
	\\[10pt]
	\caption{R-squared of the regressions of Table \ref{table:ff4_factors} with just the FF3 factors (on the left) and considering also the additional market risk factor (on the right). When adding the corresponding risk factor to the ESG promised and ESG realized regressions we observe a consistent improvement.}
	\label{table:R2}
\end{table}

In light of the presented results, we can tentatively answer the question that we have posed in the title: it is possible to hedge efficiently carbon risk. Unfortunately, the same does not hold for a standard ESG risk factor.

\section{Conclusions \& Policy implications}
\label{conclusion}
This paper introduces an innovative methodology to construct hedging portfolios to any risk factor with a focus on investigating  four sustainability-linked market risk factors. We enlighten how it is possible to build an efficient hedging portfolio within our framework for the carbon risk but not for the ESG one. The latter appears  too nebulous an  indicator to be hedged efficiently. \\

The approach combines the TMFG trimming algorithm with the node2vec embedding technique, obtaining a low-dimensional representation of the financial market correlations that also minimizes  information loss (see section \ref{methodology}). This enables the definition of efficient and explainable hedging portfolios in the embedded space.

In our empirical analysis, we employ the methodology to construct hedging portfolios of the $CO_2$, ESG, ESG promised and ESG realized factors with S\&P 500 stocks. Notably, we observe that the $CO_2$ factor is consistently close (whatever time horizon we consider) to the Utility sector, which is the most carbon intensive in the S\&P 500. Conversely, similar results are not observed for the three ESG factors (cf. Figures \ref{fig:6y}-\ref{fig:node2vecannual}).

For every risk factor, except the ESG one, we identify a statistically significant  hedging portfolio that is negatively correlated with the respective risk factor. The ESG factor appears too broadly defined to be hedged efficiently within this framework, as indicated by the absence of a statistically significant hedging portfolio, differently from the case of narrower sustainability indicators (see Tables \ref{table:ff4} and \ref{table:ff4_factors}).  Let us also emphasize that the $CO_2$ hedge portfolio  is not exposed to consistent losses in terms of standard risk metrics (cf. Table \ref{table:performance}) and is composed of 30 long-short positions in S\&P 500 stocks.

\bigskip

This study adds on the ongoing policy discourse on green/carbon capital requirements.  The inclusion of a climate related risk factor in the calculation of a bank equity capital requirements in the Basel framework can be achieved only if an appropriate hedging strategy is defined. 
From our empirical analysis, we have clear evidence that the carbon risk factor can be hedged with the proper long-short portfolio while the same does not hold for the ESG risk factor;  thus, it is possible to utilize an hedging strategy for the carbon risk factor similar to the one defined for the market risk factor. 
Consequently, the main policy implication of this study is that regulators should advocate for including the carbon risk factors in the capital requirements computation. This inclusion would also drive banks to offset their positions in carbon intensive companies with  positions in green firms and would drive portfolio decarbonization  going in the direction of the net-zero emission targets.

\section*{Acknowledgments}
The authors are grateful to T. Aste, A. Briola, D. Laurs, D. Marazzina, L. Prosperi and all participants to the International Fintech Conference 2023 in Naples and the Unwinding Complexity 2023 Conference in London for enlightening conversations on the paper  and related topics. We also thank the organizers of the Unwinding Complexity 2023 Conference for the special mention prize. The present research is part of the activities of “Dipartimento di Eccellenza 2023-2027”.

\newpage

\bibliographystyle{elsarticle-harv} 
\bibliography{sources}
\clearpage

%
\appendix
\section{Additional results}
In this appendix, we provide additional results that complement the ones presented in Section \ref{results}. In particular, i) we report the regressions estimated considering the CAPM and the FF5 models instead of the FF3 model for all risk factors and ii) the performance metrics of the \textit{close}, \textit{far} and \textit{far-close} portfolios for the three ESG-related factors.
\smallskip

In Tables \ref{table:capm} and \ref{table:ff6}, we provide an analysis similar to the one presented in Table \ref{table:ff4}, but we consider, respectively, the standard CAPM model and the FF5 model instead of the FF3 model. In both cases, we still notice that the $CO_2$ factor can be hedged with the \textit{far-close} portfolio.
\begin{table}[H]
	\scriptsize
	\centering 
	\begin{tabular}{l l l l l l l}
		\toprule
		& \multicolumn{3}{c}{\textbf{CAPM}} & \multicolumn{3}{c}{\textbf{CAPM+$CO_2$}} \\
		\cmidrule(r){2-4} \cmidrule(l){5-7}
		& \textbf{\textit{close}} & \textbf{\textit{far}} & \textbf{\textit{far-close}} & \textbf{\textit{close}} & \textbf{\textit{far}} & \textbf{\textit{far-close}} \\
		\midrule
		\rowcolor{gray!15} 
		\bm{$\alpha$} & -0.0001 & -0.0003* & -0.0002 & 0.0000 & -0.0002 & -0.0002 \\ 
		\rowcolor{gray!15} 
		& (0.0002) & (0.0001) & (0.0001) & (0.0002) & (0.0001) & (0.0001) \\ 
		\bm{$r_M-r_F$} & 0.7309*** & 1.0374*** & 0.1411*** & 0.7522*** & 1.0464*** & 0.1335*** \\ 
		& (0.0435) & (0.0261) & (0.0193) & (0.0414) & (0.0259) & (0.0184) \\ 
		\rowcolor{gray!15} 
		\bm{$CO_2$} & - & - & - & 0.3217*** & 0.137*** & -0.1146*** \\ 
		\rowcolor{gray!15} 
		& - & - & - & (0.1059) & (0.0520) & (0.0432) \\ 
		\textbf{R-squared} & 0.5379 & 0.8509 & 0.1020 & 0.5824 & 0.8573 & 0.1307 \\ 
		\bottomrule
	\end{tabular}
	\\[10pt]
	\caption{CAPM and CAPM+$CO_2$ regressions output for the \textit{close}, \textit{far} and \textit{far-close} portfolios. Statistical significance at the $10\%$, $5\%$ and $1\%$ is denoted by *, ** and ***, respectively. HAC robust standard errors are indicated in brackets.}
	\label{table:capm}
\end{table}

\begin{table}[H]
	\scriptsize
	\centering 
	\begin{tabular}{l l l l l l l}
		\toprule
		& \multicolumn{3}{c}{\textbf{Five-factor}} & \multicolumn{3}{c}{\textbf{Five-factor+$CO_2$}} \\
		\cmidrule(r){2-4} \cmidrule(l){5-7}
		& \textbf{\textit{close}} & \textbf{\textit{far}} & \textbf{\textit{far-close}} & \textbf{\textit{close}} & \textbf{\textit{far}} & \textbf{\textit{far-close}} \\
		\midrule
		\rowcolor{gray!15} 
		\bm{$\alpha$} & -0.0001 & -0.0001 & -0.0000 & -0.0000 & -0.0001 & -0.0001 \\  
		\rowcolor{gray!15} 
		& (0.0002) & (0.0001) & (0.0001) & (0.0002) & (0.0001) & (0.0001)\\
		\bm{$r_M-r_F$} & 0.7908*** & 1.0358*** & 0.1289*** & 0.8156*** & 1.0402*** & 0.1178*** \\  
		& (0.0427) & (0.0300) & (0.0220) & (0.0427) & (0.0301) & (0.0221)\\ 
		\rowcolor{gray!15} 
		\bm{$SMB$} & -0.1228** & 0.2927*** & 0.2260*** & -0.1714*** & 0.2840*** & 0.2478*** \\ 
		\rowcolor{gray!15} 
		& (0.0484) & (0.0490) & (0.0269) & (0.0501) & (0.0478) & (0.0288)\\ 
		\bm{$HML$} & 0.0287 & 0.1762*** & 0.0617* & -0.0573 & 0.1608*** & 0.1004*** \\ 
		& (0.0449) & (0.0408) & (0.0333) & (0.0538) & (0.0428) & (0.0368)\\ 
		\rowcolor{gray!15} 
		\bm{$RMW$} & 0.2062*** & 0.4287*** & 0.1458*** & 0.1202* & 0.4132*** & 0.1844*** \\ 
		\rowcolor{gray!15} 
		& (0.0707) & (0.0466) & (0.0418) & (0.0723) & (0.0482) & (0.0441)\\  
		\bm{$CMA$} & 0.4481*** & -0.2216** & -0.3749*** & 0.2916** & -0.2497*** & -0.3047*** \\  
		& (0.0959) & (0.0877) & (0.0714) & (0.1245) & (0.0905) & (0.0790)\\  
		\rowcolor{gray!15} 
		\bm{$CO_2$} & - & - & - & 0.3247** & 0.0584 & -0.1458** \\ 
		\rowcolor{gray!15} 
		& - & - & -  & (0.1394) & (0.0412) & (0.0651)\\ 
		\textbf{R-squared} & 0.5936 & 0.8736 & 0.2415 & 0.6256 & 0.8743 & 0.2700 \\     
		\bottomrule
	\end{tabular}
	\\[10pt]
	\caption{Fama-French five-factor and five-factor+$CO_2$ regressions output for the \textit{close}, \textit{far} and \textit{far-close} portfolios. Statistical significance at the $10\%$, $5\%$ and $1\%$ is denoted by *, ** and ***, respectively. HAC robust standard errors are indicated in brackets.}
	\label{table:ff6}
\end{table}

Tables \ref{table:capm_factors} and \ref{table:ff6_factors} display the results of the modified CAPM and FF5 models for the ESG, ESG promised and ESG realized factors. Similarly to the case of the FF3 model in Table \ref{table:ff4_factors}, these results show that only the ESG promised and realized factors can be identified as significant risk factors. Moreover, when we examine the portfolio performance measures reported in Table \ref{table:performance_factors}, we notice that the \textit{far-close} portfolios offer an efficient hedging opportunity, as in the case of the CO2 risk factor. This is evident from their considerably lower MDD and VaR compared to the other alternatives.

\begin{table}[H]
	
	\centering 
	\resizebox{\textwidth}{!}{  \begin{tabular}{l l l l l l l l l l}
			\toprule
			& \multicolumn{3}{c}{\textbf{CAPM+ESG}} & \multicolumn{3}{c}{\textbf{CAPM+ESGp}} & \multicolumn{3}{c}{\textbf{CAPM+ESGr}}\\
			\cmidrule(r){2-4} \cmidrule(l){5-7} \cmidrule(l){8-10}
			& \textbf{\textit{close}} & \textbf{\textit{far}} & \textbf{\textit{far-close}} & \textbf{\textit{close}} & \textbf{\textit{far}} & \textbf{\textit{far-close}} & \textbf{\textit{close}} & \textbf{\textit{far}} & \textbf{\textit{far-close}}\\
			\midrule
			\rowcolor{gray!15} 
			\bm{$\alpha$} & -0.0000 & -0.0002 & 0.0000 & 0.0002** & -0.0003 & -0.0003** & -0.0002** & -0.0002 & 0.0001 \\  
			\rowcolor{gray!15}
			& (0.0001) & (0.0002) & (0.0001) & (0.0001) & (0.0002) & (0.0001) & (0.0001) & (0.0002) & (0.0001) \\ 
			\bm{$r_M-r_F$} & 0.9566*** & 1.0661*** & 0.0727** & 0.9357*** & 1.0865*** & 0.0578* & 1.0635*** & 1.0444*** & -0.0209 \\ 
			& (0.0167) & (0.0449) & (0.0321) & (0.0151) & (0.0436) & (0.0326) & (0.0160) & (0.0439) & (0.0299) \\  
			\rowcolor{gray!15}
			\bm{$factor$} & -0.1322*** & 0.0706 & 0.0417 & 0.0370* & -0.0858 & -0.0774** & 0.0425*** & -0.1417* & -0.0954** \\ 
			\rowcolor{gray!15}
			& (0.0376) & (0.0718) & (0.0364) & (0.0193) & (0.0526) & (0.0314) & (0.0139) & (0.0739) & (0.0453) \\ 
			\textbf{R-squared} & 0.8960 & 0.7827 & 0.0461 & 0.8478 & 0.7750 & 0.0339 & 0.8888 & 0.7848 & 0.0274 \\ 
			\bottomrule
	\end{tabular}}
	\\[10pt]
	\caption{CAPM+ESG/ESGp/ESGr regressions output for the \textit{close}, \textit{far} and \textit{far-close} portfolios. Statistical significance at the $10\%$, $5\%$ and $1\%$ is denoted by *, ** and ***, respectively. HAC robust standard errors are indicated in brackets.}
	\label{table:capm_factors}
\end{table}

\begin{sidewaystable} 
	\begin{table}[H]
		\small
		\centering 
		\begin{tabular}{l l l l l l l l l l}
			\toprule
			& \multicolumn{3}{c}{\textbf{Five-factor+ESG}} & \multicolumn{3}{c}{\textbf{Five-factor+ESGp}} & \multicolumn{3}{c}{\textbf{Five-factor+ESGr}}\\
			\cmidrule(r){2-4} \cmidrule(l){5-7} \cmidrule(l){8-10}
			& \textbf{\textit{close}} & \textbf{\textit{far}} & \textbf{\textit{far-close}} & \textbf{\textit{close}} & \textbf{\textit{far}} & \textbf{\textit{far-close}} & \textbf{\textit{close}} & \textbf{\textit{far}} & \textbf{\textit{far-close}}\\
			\midrule
			\rowcolor{gray!15}   
			\bm{$\alpha$} & -0.0001 & -0.0001 & 0.0001 & 0.0001 & -0.0001 & -0.0002* & -0.0000 & -0.0003 & -0.0000 \\ 
			\rowcolor{gray!15}
			& (0.0001) & (0.0002) & (0.0001) & (0.0001) & (0.0001) & (0.0001) & (0.0001) & (0.0002) & (0.0001) \\ 
			\bm{$r_M-r_F$} & 0.9419*** & 1.0298*** & 0.0722*** & 0.9643*** & 0.9945*** & 0.0495** & 1.0050*** & 1.0416*** & 0.0047 \\ 
			& (0.0136) & (0.0365) & (0.0234) & (0.0160) & (0.0307) & (0.0218) & (0.0129) & (0.0424) & (0.0272)\\ 
			\rowcolor{gray!15}
			\bm{$SMB$} & 0.1332*** & 0.2139*** & 0.0741 & 0.1138*** & 0.1559*** & 0.0129 & 0.0946*** & 0.2970*** & 0.0924* \\   
			\rowcolor{gray!15}
			& (0.0248) & (0.0611) & (0.0572) & (0.0311) & (0.0407) & (0.0272) & (0.0264) & (0.0746) & (0.0547)\\  
			\bm{$HML$} & -0.2022*** & 0.4509*** & 0.3110*** & -0.2819*** & 0.3325*** & 0.2991*** & 0.0844*** & 0.4593*** & 0.1767*** \\  
			& (0.0205) & (0.0513) & (0.0349) & (0.0241) & (0.0374) & (0.0222) & (0.0209) & (0.0616) & (0.0425)\\ 
			\rowcolor{gray!15}
			\bm{$RMW$} & -0.0869*** & 0.3516*** & 0.1550*** & -0.1666*** & 0.3408*** & 0.3264*** & -0.1485*** & 0.3862*** & 0.2922*** \\ 
			\rowcolor{gray!15}
			& (0.0272) & (0.0502) & (0.0357) & (0.0353) & (0.0438) & (0.0305) & (0.0333) & (0.0585) & (0.0432)\\ 
			\bm{$CMA$} & -0.0541 & -0.2951*** & -0.1760** & -0.1328*** & -0.0120 & 0.0319 & -0.1820*** & -0.2905** & -0.0802 \\
			& (0.0384) & (0.1004) & (0.0805) & (0.0485) & (0.0672) & (0.0462) & (0.0452) & (0.1211) & (0.0832)\\ 
			\rowcolor{gray!15}
			\bm{$factor$} & -0.0310** & -0.0826** & -0.0176 & 0.0520** & -0.0589 & -0.0668*** & 0.0309** & -0.1385* & -0.0832** \\ 
			\rowcolor{gray!15}
			& (0.0157) & (0.0323) & (0.0290) & (0.022) & (0.0386) & (0.0193) & (0.0147) & (0.0736) & (0.0420)\\  
			\textbf{R-squared} & 0.9146 & 0.8450 & 0.3358 & 0.8885 & 0.8705 & 0.4260 & 0.9187 & 0.8106 & 0.1676 \\   
			\bottomrule
		\end{tabular}
		\\[10pt]
		\caption{Fama-French five-factor+ESG/ESGp/ESGr regressions output for the \textit{close}, \textit{far} and \textit{far-close} portfolios. Statistical significance at the $10\%$, $5\%$ and $1\%$ is denoted by *, ** and ***, respectively. HAC robust standard errors are indicated in brackets.}
		\label{table:ff6_factors}
	\end{table}
\end{sidewaystable}
\begin{table}[H]
	\centering 
	\resizebox{\textwidth}{!}{\begin{tabular}{c c c c c c c c c c}
			\toprule
			& \multicolumn{3}{c}{\textbf{ESG}} & \multicolumn{3}{c}{\textbf{ESGp}} & \multicolumn{3}{c}{\textbf{ESGr}}\\
			\cmidrule(r){2-4} \cmidrule(l){5-7} \cmidrule(l){8-10}
			& \textbf{\textit{close}} & \textbf{\textit{far}} & \textbf{\textit{far-close}} & \textbf{\textit{close}} & \textbf{\textit{far}} & \textbf{\textit{far-close}} & \textbf{\textit{close}} & \textbf{\textit{far}} & \textbf{\textit{far-close}}\\
			\midrule
			\rowcolor{gray!15}
			\textbf{Sharpe ratio} & 0.627 & 0.316 & 0.100 & 0.863 & 0.355 & -0.488 & 0.634 & 0.215 & -0.138 \\ 
			\textbf{Sortino ratio} & 0.743 & 0.322 & 0.115 & 1.053 & 0.367 & -0.659 & 0.735 & 0.216 & -0.163 \\ 
			\rowcolor{gray!15}
			\textbf{Omega ratio} & 1.137 & 1.087 & 1.048 & 1.182 & 1.095 & 0.928 & 1.139 & 1.062 & 0.993 \\ 
			\textbf{MDD} & 31.3\% & 54.9\% & 28.1\% & 29.8\% & 50.3\% & 28.4\% & 36.9\% & 59.9\% & 27.4\% \\ 
			\rowcolor{gray!15}
			\textbf{VaR 5\%} & 1.9\% & 2.4\% & 0.8\% & 2.0\% & 2.3\% & 0.9\% & 2.1\% & 2.6\% & 0.9\% \\ 
			\bottomrule
	\end{tabular}}
	\\[10pt]
	\caption{Evaluation measures for the \textit{close}, \textit{far} and \textit{far-close} portfolios constructed according to the ESG, ESGp and ESGr factors.}
	\label{table:performance_factors}
\end{table}

\end{document}